\documentclass[%
reprint,
showpacs,preprintnumbers,
 amsmath,amssymb,
 aps,
]{revtex4-1}

\usepackage{graphicx}
\usepackage{dcolumn}
\usepackage{bm}
\usepackage{subfigure} 
\usepackage{color,soul}
\usepackage{ulem}
\usepackage{booktabs}

\begin{document}


\title{Effect of line tension on axisymmetric nanoscale capillary bridges
at the liquid-vapor equilibrium 
}

\author{Masao Iwamatsu}
\altaffiliation[Permanent address ]{Tokyo City University, Setagaya-ku, Tokyo 158-8557, Japan}
\email{iwamatm@tcu.ac.jp (corresponding author)}
\author{Hiroyuki Mori}%
\affiliation{
Department of Physics, Tokyo Metropolitan University, Hachioji, Tokyo 192-0397, Japan
}%



\date{\today}

\begin{abstract}
The effect of line tension on the axisymmetric nanoscale capillary bridge between two identical substrates with convex, concave and flat geometry at the liquid-vapor equilibrium is theoretically studied.  The modified Young's equation for the contact angle, which takes into account the effect of line tension is derived on a general axisymmetric curved surface using the variational method.  Even without the effect of line tension, the parameter space where the bridge can exist is limited simply by the geometry of substrates.  The modified Young's equation further restricts the space where the bridge can exist when the line tension is positive because the equilibrium contact angle always remain finite and the wetting state near the zero contact angle cannot be realized. It is shown that the interplay of the geometry and the positive line tension restricts the formation of capillary bridge.
\end{abstract}

\maketitle

\section{Introduction}
The nanoscale liquid bridge which forms in a narrow slit between two small objects is a ubiquitous phenomenon occurring in humid atmospheres in our daily life.  It plays a fundamental role in many natural and geological phenomena such as the water retention in soil~\cite{Haines1925,Fisher1927,Leelamanie2012,Duriez2017}, the wet friction~\cite{Lee2014}, the biological adhesion of some insects~\cite{Federle2002,Su2009}, and the cloud formation~\cite{Crouzet1995} etc.  It also plays an important role in various industrial processing such as that of the powders and the granular matters~\cite{Bocquet1998} and, in particular, in the modern nano-technologies and nano-fluidics~\cite{Rozynek2017,Pi2018}.

This liquid bridge called capillary bridge forms via the heterogeneous nucleation of capillary condensation of volatile liquids~\cite{Evans1986,Iwamatsu1996,Restagno2000,Israelachivili2011,Desgranges2017,Kim2018}.  This capillary condensation is a special case of the heterogeneous nucleation in which the condensation can occur at the liquid-vapor equilibrium, and even in an undersaturated vapor with a relative humidity lower than the saturation.  The capillary bridge corresponds to the critical nucleus of capillary condensation.  In contrast, the usual homogeneous nucleation of a spherical droplet can occur only in an oversaturated vapor with relative humidity higher than the saturation.  Then, the capillary bridge can easily form, which is the reason why the capillary condensation is so ubiquitous.  Therefore, various theoretical~\cite{Haines1925,Fisher1927,Pepin2000,Kruyt2017}, numerical~\cite{Erle1971,Orr1975,Lian2016}, and experimental~\cite{Butt2009,Kwon2018,Kim2018} studies have been conducted by many engineers and scientists for many years.

After the invention of various force spectrosopy techniques such as the surface force apparatus (SFA) and the atomic force microscopes (AFM), the problem of capillary bridge, in particular, the normal capillary force by the bridge have been attracted renewed interest recently~\cite{Butt2009,Asay2010,Crassous2011}.  Since the nanoscale liquid bridges can easily form even in the low relative humidity, it is also a convenient tool to study the surface properties such as the size-dependence of surface tension of the nanoscale liquid bridges~\cite{Kwon2018,Kim2018}.  It can also serve as a testing ground of various statistical mechanical theory of nanoscale liquids~\cite{Dobbs1992a,Bauer2000,Malijevsky2015,MacDowell2018,Malijevsky2019}.  

Most of those theoretical studies rely on the atomic simulations~\cite{Giovambattista2016,Desgranges2017}, the mesoscopic disjoining pressure, and the density functional theory using various model surface potentials~\cite{Evans1986,Iwamatsu1996,Malijevsky2015,MacDowell2018}.  However, the classical capillary theory using the macroscopic concepts such as the surface tension~\cite{Rowlinson1982,Israelachivili2011} and the line tension~\cite{Rowlinson1982,Gibbs1906} is still useful~\cite{Malijevsky2015,Iwamatsu2015a,Iwamatsu2015b}.  In particular, the analytical or semi-analytical results obtained from the classical capillary theory have been useful to understand the physics of capillary phenomena and to analyze experimental results directly~\cite{Kwon2018,Kim2018}.

In fact, Malijevsk\'y and Parry~\cite{Malijevsky2018} recently showed that the prediction of capillary condensation from the classical capillary theory remains highly accurate
down to the order of tens of molecular diameters.  They reached this conclusion by comparing the results from the classical capillary theory with those from the non-local density functional theory~\cite{Malijevsky2015,Malijevsky2019}, which is more accurate than the local square-gradient density functional theory~\cite{Evans1979} or the second gradient theory~\cite{Dellisola1995}.  There are growing evidences that the classical capillary theory is accurate down to the order of nanometer and that it can be served as a minimal model of nanoscale liquids~\cite{Malijevsky2015,Malijevsky2019}. The effect of disjoining pressure or the surface potential, for example, can be partly taken into account by the line tension~\cite{Indekeu1994,Aveyard1999}.

In the present study, we study the axisymmetric nanoscale capillary bridge at the liquid-vapor equilibrium theoretically using the classical capillary theory, which is believed to be accurate down to nanometer scale~\cite{Malijevsky2015, Giovambattista2016,Kim2018,Kwon2018}.  We consider the axisymmetric bridge because the surface with constant contact angle must be rotationally symmetric (axisymmetric)~\cite{Wente1980,Finn1986}. We also consider the bridge only at the liquid-vapor equilibrium because the liquid-vapor meniscus is exactly given by the catenary~\cite{Orr1975,Malijevsky2015,Eriksson1995}, and we can avoid numerical uncertainty of the shape of meniscus due to the numerical integration of differential equations. The microscopic effects for the nanoscale bridges will be taken into account by considering the effective line tension~\cite{Gibbs1906,Navascues1981,Widom1995,Schimmele2007,Hienola2007,Iwamatsu2015a,Law2017}.  This strategy has been already adopted by Dutka and Napi\'orkowski~\cite{Dutka2007,Dutka2010}, and Aveyard et al.~\cite{Aveyard1999} who showed that the effective line tension can include the effect of disjoining pressure.  However, they considered only special geometries such as the AFM-like geometry.  In this paper, we consider a few typical geometries~\cite{Butt2009,Tselishchev2003}.  Also, we pay most attention to the formation of bridge under the influence of a positive line tension.  Since we will not consider the microscopic disjoining pressure or the surface potential, we will not consider the detail of surface phase transition such as the wetting and the prewetting transition~\cite{Dobbs1992a,Bauer2000,Malijevsky2015}.

This paper is organized as follows. In Sec. II, we present a general modified Young's equation, which takes into account the line tension, to determine the equilibrium contact angle on a axisymmetric curved substrate using the variational approach~\cite{Hildebrand1992,Bormashenko2009}.  In Sec. III, we use the modified Young's equation to study the effect of line tension on capillary bridges confined in slits of several typical geometries.  We will show that the positive line tension will severely restrict the parameter space where the capillary bridge can exist even thought the line tension does not contribute directly to the normal capillary force~\cite{Butt2009}.   Finally, in Sec. IV, we conclude by emphasizing the implication of our results to future experimental, numerical and theoretical studies of nanoscale bridges.

\section{\label{sec:sec2} Generalized Young's equation on axisymmetric curved substrates}

\subsection{Convex substrate}

\begin{figure}[htbp]
\begin{center}
\includegraphics[width=0.6\linewidth]{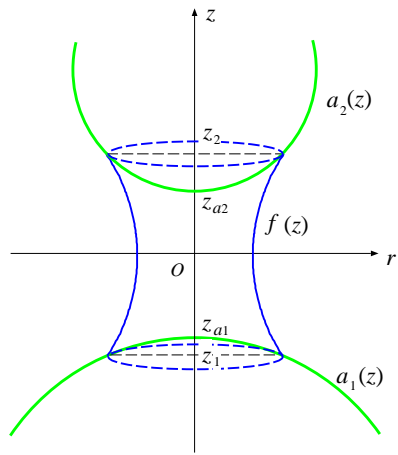}
\caption{
The capillary bridge $f(z)$ connecting two convex substrate $a_{1}(z)$ and $a_{2}(z)$ at $z_{1}$ and $z_{2}$.
 }
\label{fig:br1}
\end{center}
\end{figure}

We consider the effect of line tension on the contact angle of liquid bridge called capillary bridge shown in Fig.~\ref{fig:br1} when the three-phase contact line is on an axisymmetric curved substrate.  We assume that the geometry is axisymmetric around $z-$axis~\cite{Butt2009,Tselishchev2003}.  We use the classical capillary theory so that the liquid-vapor surface is assumed to be sharp and the surface tension is constant and does not depend on the curvature of the meniscus.  Then, the meniscus of the liquid bridge is represented by a function $f(z)$.  Similarly, the two substrates which are connected by the liquid bridge are represented by two functions $a_{1}(z)$ and $a_{2}(z)$ as shown in Fig.~\ref{fig:br1}. They are connected by the liquid bridge spanning between $z_{1}$ and $z_{2}$.

Since we consider the capillary bridge of capillary condensation, we use the free energy called the grand potential~\cite{Dutka2007,Dutka2010,Iwamatsu2015a}.  The total free energy $\Delta\Omega\left[f\right]$ of the system in the capillary theory consists of three contributions
\begin{equation}
\Delta\tilde{\Omega}=\frac{\Delta\Omega\left[f\right]}{2\pi\sigma_{\rm lv}}=\Delta\tilde{\Omega}_{\rm lv}+\Delta\tilde{\Omega}_{\rm sl}+\Delta\tilde{\Omega}_{\rm slv},
\label{eq:Br1}
\end{equation}
where the free energy $\Delta\Omega$ is divided by $2\pi\sigma_{\rm lv}$, where $\sigma_{\rm lv}$ is the liquid-vapor surface tension, and 
\begin{equation}
\Delta\tilde{\Omega}_{\rm lv}\left[f\right] 
= \int_{z_{1}}^{z_{2}} dz F\left[z;f,f'\right] 
\label{eq:Br2}
\end{equation}
is the free energy of the liquid bridge, and
\begin{eqnarray}
F\left[z,f,f'\right] &=& f\left(z\right)\sqrt{1+f'\left(z\right)^{2}} \nonumber \\
&& +\Delta\tilde{p}\left( f\left(z\right)^2-a_{1}\left(z\right)^2-a_{2}\left(z\right)^2  \right)
\label{eq:Br3}
\end{eqnarray}
is the free-energy density of the bridge (liquid), where $\Delta\tilde{p}=\Delta p/\sigma_{\rm lv}$ represents the pressure difference $\Delta p$ between the liquid bridge and the surrounding vapor.  Equation (\ref{eq:Br3}) consists of the surface energy (the first term) and the volume energy (the second term).  The prime means the spacial derivative $f'=df/dz$.  In order to calculate the volume energy, we should take into account the appropriate boundary conditions by the geometry of substrates.

The second term of Eq.~(\ref{eq:Br1}) is the solid-liquid surface energy due to the wetting of the solid substrate by the liquid bridge given by
\begin{eqnarray}
\Delta\tilde{\Omega}_{\rm sl}\left[f\right] &=& 
-\int_{z_{1}}^{z_{a1}}dz \cos\theta_{\rm Y1} a_{1}\left(z\right)\sqrt{1+a_{1}'\left(z\right)^2} \nonumber \\
&&-\int_{z_{a2}}^{z_{2}}dz \cos\theta_{\rm Y2} a_{2}\left(z\right)\sqrt{1+a_{2}'\left(z\right)^2}, 
\label{eq:Br4}
\end{eqnarray}
which consists of the contribution from upper (2) and lower (1) substrates, where $z_{a1}$ is the top of the substrate $a_{1}$ and $z_{a2}$ is the bottom of the substrate $a_{2}$.  The solid-liquid surface energies in Eq.~(\ref{eq:Br4}) are characterized by the wettability of substrates represented by two Young's contact angle $\theta_{\rm Y1}$ and $\theta_{\rm Y2}$ defined by
\begin{eqnarray}
\cos\theta_{\rm Y1}&=&\frac{\left(\sigma_{\rm sv}\right)_{1}-\left(\sigma_{\rm sl}\right)_{1}}{\sigma_{\rm lv}},
\nonumber \\
\cos\theta_{\rm Y2}&=&\frac{\left(\sigma_{\rm sv}\right)_{2}-\left(\sigma_{\rm sl}\right)_{2}}{\sigma_{\rm lv}},
\label{eq:Br5}
\end{eqnarray}
where $\sigma_{\rm sv}$ and $\sigma_{\rm sl}$ are the solid-vapor and solid-liquid surface energy, and the subscripts 1 and 2 refer to the two substrates 1 and 2.  

The last term of Eq.~(\ref{eq:Br1}) is the line-tension energy at the two three-phase (solid-liquid-vapor) contact lines given by
\begin{eqnarray}
\Delta\tilde{\Omega}_{\rm slv} 
&=& \tilde{\tau}_{1}a_{1}\left(z_{1}\right) +  \tilde{\tau}_{2}a_{2}\left(z_{2}\right),
\label{eq:Br6}
\end{eqnarray}
where the line tensions $\tau_{1}$ and $\tau_{2}$ are scaled by $\sigma_{\rm lv}$ as $\tilde{\tau}_{1}=\tau_{1}/\sigma_{\rm lv}$ and  $\tilde{\tau}_{2}=\tau_{2}/\sigma_{\rm lv}$.

\begin{figure}[htbp]
\begin{center}
\includegraphics[width=0.8\linewidth]{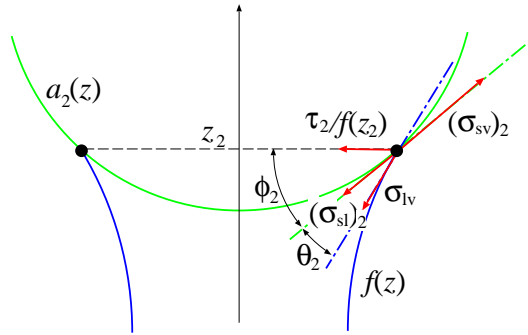}
\caption{
The contact angle $\theta_{2}$ between the tangent to the substrate $a_{2}(z)$ and that to the meniscus $f(z)$. The line tension force $\tau_{2}/f\left(z_{2}\right)$ which lies on the circular plane made from the three-phase contact line is always perpendicular to the rotational axis $z$. Therefore, it does not contribute to the normal capillary force.
 }
\label{fig:Br2}
\end{center}
\end{figure}

The meniscus $f(z)$ of the capillary bridge is determined from the variation of the total free energy.  The boundaries of integration $z_{1}$ and $z_{2}$ are not fixed.  Rather, the meniscus $f(z)$ are required to touch the surface of the substrate $a_{1}(z)$ and $a_{2}(z)$.  Therefore, the boundary conditions of the variational problem are
\begin{eqnarray}
f\left(z_{1}\right)&=&a_{1}\left(z_{1}\right), \nonumber \\
f\left(z_{2}\right)&=&a_{2}\left(z_{2}\right).
\label{eq:Br7}
\end{eqnarray}
Using the standard variation~\cite{Hildebrand1992} by $\delta f$ as well as by the variable end points $\delta z_{1}$ and $\delta z_{2}$ (Appendix), we can derive the Euler-Lagrange equation
\begin{equation}
\frac{\partial F}{\partial f}-\frac{d}{dz}\left(\frac{\partial F}{\partial f'}\right)=2\Delta\tilde{p},
\label{eq:Br8}
\end{equation}
which is written explicitly as~\cite{Dutka2007,Dutka2010}
\begin{equation}
\frac{1}{f\left( z\right)\sqrt{1+f'(z)^{2}}}-\frac{d}{dz}\frac{f'(z)}{\sqrt{1+f'(z)^{2}}}=-2\Delta\tilde{p},
\label{eq:Br9}
\end{equation}
and determines the capillary meniscus $f\left(z\right)$.  Since we will consider the liquid-vapor equilibrium, we will set $\Delta\tilde{p}=0$ later in the next section.  In addition, the boundary condition at $z_{2}$, for example, called {\it transversality condition}~\cite{Hildebrand1992,Bormashenko2009} is given by (see Appendix for detail)
\begin{eqnarray}
\cos\theta_{\rm Y2}
&-& \frac{1+f'\left(z_{2}\right)a_{2}'\left(z_{2}\right)}{\sqrt{1+f'\left(z_{2}\right)^{2}}\sqrt{1+a_{2}'\left(z_{2}\right)^{2}}}
\nonumber \\
&&-\tilde{\tau}_{2}\frac{a_{2}'\left(z_{2}\right)}{a_{2}\left(z_{2}\right)\sqrt{1+a_{2}'\left(z_{2}\right)^{2}}}=0,
\label{eq:Br10}
\end{eqnarray}
where the second term is the cosine of the equilibrium contact angle $\theta_{2}$ at $z_{2}$ (Fig. \ref{fig:Br2})
\begin{equation}
\cos\theta_{2}=\frac{1+f'\left(z_{2}\right)a_{2}'\left(z_{2}\right)}{\sqrt{1+f'\left(z_{2}\right)^{2}}\sqrt{1+a_{2}'\left(z_{2}\right)^{2}}}
\label{eq:Br11}
\end{equation}
between the tangent to the substrate $a_{2}(z)$ and that to the liquid-vapor surface $f(z)$.  The boundary condition at $z_{1}$ will be obtained simply by replacing the subscript 2 by 1.  Equations (\ref{eq:Br10}) and (\ref{eq:Br11}) lead to
\begin{equation}
\cos\theta_{2}=\cos\theta_{\rm Y2}-\tilde{\tau}_{2}\frac{a_{2}'\left(z_{2}\right)}{a_{2}\left(z_{2}\right)\sqrt{1+a_{2}'\left(z_{2}\right)^{2}}},
\label{eq:Br12}
\end{equation}
which is the general form of the modified Young's equation on an axisymmetric convex substrate with $a_{2}'\left(z_{2}\right)>0$.

Since the angle $\phi_{2}$ between the tangent to $a_{2}\left(z\right)$ at $z_{2}$ and the three-phase contact line perpendicular to the rotational axis ($z$-axis) satisfies (Fig.~\ref{fig:Br2})
\begin{equation}
\cos\phi_{2}=\frac{a_{2}'\left(z_{2}\right)}{\sqrt{1+a_{2}'\left(z_{2}\right)^{2}}},
\label{eq:Br13}
\end{equation}
Eq. (\ref{eq:Br12}) is written as
\begin{equation}
\cos \theta_{2}=\cos\theta_{\rm Y2}
-\frac{\tilde{\tau}_{2}}{a_{2}\left(z_{2}\right)}\cos\phi_{2},
\label{eq:Br14}
\end{equation}
or
\begin{equation}
\cos \theta_{2}=\cos\theta_{\rm Y2}
-\frac{\tilde{\tau}_{2}}{f\left(z_{2}\right)}\cos\phi_{2},
\label{eq:Br15}
\end{equation}
which is also know as the modified Young's equation.  The effect of line tension $\tau_{2}$ on the equilibrium contact angle $\theta_{2}$ on an axisymmetric surface is determined from Eq.~(\ref{eq:Br15}).  It also expresses the force balance condition of the four tensions due to the three surface tensions $\sigma_{\rm lv}$, $\left(\sigma_{\rm sl}\right)_{2}$, and $\left(\sigma_{\rm sv}\right)_{2}$, and that due to the line tension $\tau_{\rm 2}/f\left(z_{2}\right)$ along the tangent to the surface $a_{2}\left(z\right)$ at $z_{2}$ (Fig.~\ref{fig:Br2}), which is the generalization of a similar force balance condition on a spherical surface~\cite{Hienola2007,Iwamatsu2015b}.  A boundary condition on a conical surface has already been obtained by Dukta and Napi\'orkovsky~\cite{Dutka2007,Dutka2010}.

When the substrate is convex ($\cos\phi_{2}>0$ in Eq.~(\ref{eq:Br15}) or $a_{2}'\left( z_{2}\right)>0$ in Eq.~(\ref{eq:Br12})) and hydrophilic ($\cos\theta_{\rm Y2}>0$), a positive line tension ($\tilde{\tau}_{2}>0$) makes the equilibrium contact angle $\theta_{2}$ larger than Young's contact angle $\theta_{\rm Y2}$ since $0<\cos\theta<\cos\theta_{\rm Y2}$.  Therefore, a positive line tension makes the hydrophilic substrate less hydrophilic ($\theta>\theta_{\rm Y2}$).

Since the line-tension force  $\tau_{2}/f\left(z_{2}\right)$ acting at the three-phase contact line always stays within the circular plane of the three-phase contact line (Fig.~\ref{fig:Br2}), the line tension does not contribute directly to the normal capillary force~\cite{Butt2009,Kwon2018,Kim2018} parallel to the rotational axis ($z-$axis).  Therefore, the normal capillary force can probe only the surface-tension force at the three phase contact line and the capillary pressure due to the pressure difference between the liquid and the vapor phase.  The line tension affects the normal capillary force and the condensation indirectly through the modification of contact angle~\cite{Duriez2017} through Eq.~(\ref{eq:Br10}) or (\ref{eq:Br15}).

\subsection{Concave substrate}

\begin{figure}[htbp]
\begin{center}
\includegraphics[width=0.6\linewidth]{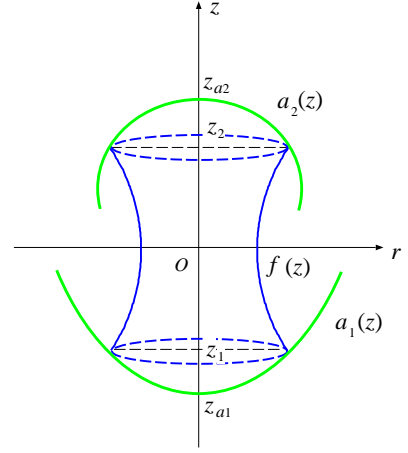}
\caption{
The capillary bridge $f(z)$ connecting two concave substrates $a_{1}(z)$ and $a_{2}(z)$ at $z_{1}$ and $z_{2}$.
 }
\label{fig:Br3}
\end{center}
\end{figure}

For the concave substrate shown in Fig.~\ref{fig:Br3}, the upper and the lower bounds of the integral should be exchanged in Eq.~(\ref{eq:Br4}), which is now written as
\begin{eqnarray}
\Delta\tilde{\Omega}_{\rm sl}\left[f\right] &=& 
-\int_{z_{a1}}^{z_{1}}dz \cos\theta_{\rm Y1} a_{1}\left(z\right)\sqrt{1+a_{1}'\left(z\right)^2} \nonumber \\
&&-\int_{z_{2}}^{z_{a2}}dz \cos\theta_{\rm Y2} a_{2}\left(z\right)\sqrt{1+a_{2}'\left(z\right)^2},
\label{eq:Br16}
\end{eqnarray}
where $z_{a1}$ is the bottom of the substrate $a_{1}$ and $z_{a2}$ is the top of the substrate $a_{2}$.  Then, the transversality condition in Eq.~(\ref{eq:Br10}) should be replaced by
\begin{eqnarray}
-\cos\theta_{\rm Y2}
&-& \frac{1+f'\left(z_{2}\right)a_{2}'\left(z_{2}\right)}{\sqrt{1+f'\left(z_{2}\right)^{2}}\sqrt{1+a_{2}'\left(z_{2}\right)^{2}}}
\nonumber \\
&&-\tilde{\tau}_{2}\frac{a_{2}'\left(z_{2}\right)}{a_{2}\left(z_{2}\right)\sqrt{1+a_{2}'\left(z_{2}\right)^{2}}}=0,
\label{eq:Br17}
\end{eqnarray}
where the second term is related to the cosine of the equilibrium contact angle $\theta_{2}$ at $z_{2}$ through
\begin{equation}
\cos\theta_{2}=-\frac{1+f'\left(z_{2}\right)a_{2}'\left(z_{2}\right)}{\sqrt{1+f'\left(z_{2}\right)^{2}}\sqrt{1+a_{2}'\left(z_{2}\right)^{2}}}
\label{eq:Br18}
\end{equation}
for the concave substrate. Equations (\ref{eq:Br17}) and (\ref{eq:Br18}) lead to
\begin{equation}
\cos\theta_{2}=\cos\theta_{\rm Y2}+\tilde{\tau}_{2}\frac{a_{2}'\left(z_{2}\right)}{a_{2}\left(z_{2}\right)\sqrt{1+a_{2}'\left(z_{2}\right)^{2}}},
\label{eq:Br19}
\end{equation}
which is the general form of the modified Young's equation on an axisymmetric concave substrate with $a_{2}'\left(z_{2}\right)<0$, which corresponds to Eq.~(\ref{eq:Br12}) for a convex substrate with $a_{2}'\left(z_{2}\right)>0$.

Therefore, a positive line tension ($\tilde{\tau}_{2}>0$) also makes the equilibrium contact angle $\theta_{2}$ larger than Young's contact angle $\theta_{\rm Y2}$ ($0<\cos\theta_{2}<\cos\theta_{\rm Y2}$) on a hydrophilic substrate from Eq.~(\ref{eq:Br19}) since $a_{2}'\left( z_{2}\right)<0$ on a concave substrate.

\subsection{Flat substrate}

Although the effect of line tension on a flat substrate is already well documented~\cite{Navascues1981,Widom1995}, we briefly touch the case when the substrate is horizontally flat and infinite.  Suppose the substrate $a_{1}(z)$ at $z_{1}$ is flat, the grand potential is given by
\begin{eqnarray}
\Delta\tilde{\Omega}\left[f\right]
&=& \int_{z_{1}}^{z_{2}} dz F\left[z,f,f'\right] \nonumber \\
&-&\int_{z_{1}}^{z_{2}}dz \cos\theta_{\rm Y2} a_{2}\left(z\right)\sqrt{1+a_{2}'\left(z\right)^2} \nonumber \\
&-& \cos\theta_{\rm Y1}\frac{f\left(z_{1}\right)^{2}}{2} +\tilde{\tau}_{1}f\left(z_{1}\right) +  \tilde{\tau}_{2}a_{2}\left(z_{2}\right).
\label{eq:Br20}
\end{eqnarray}
Since the boundary $z_{1}$ is fixed and $f\left(z_{1}\right)$ is unknown, we have to impose the {\it natural boundary condition}~\cite{Hildebrand1992} instead of the transversality condition in Eq.~(\ref{eq:Br10}).  The natural boundary condition at $z_{1}$ for the free energy in Eq.~(\ref{eq:Br20}) is given by~\cite{Dutka2007,Dutka2010}
\begin{equation}
\cos\theta_{\rm Y1}+\frac{f'\left(z_{1}\right)}{\sqrt{1+f'\left(z_{1}\right)^{2}}}-\frac{\tilde{\tau}_{1}}{f\left(z_{1}\right)}=0,
\label{eq:Br21}
\end{equation}
where the equilibrium contact angle $\theta_{1}$ at $z_{1}$ is given by 
\begin{equation}
\cos\theta_{1}=\frac{f'\left(z_{1}\right)}{\sqrt{1+f'\left(z_{1}\right)^{2}}},
\label{eq:Br22}
\end{equation}
and the boundary condition Eq.~(\ref{eq:Br21}) becomes
\begin{equation}
\cos\theta_{1}=\cos\theta_{\rm Y1}-\frac{\tilde{\tau}_{1}}{f\left(z_{1}\right)},
\label{eq:Br23}
\end{equation}
which is the modified Young's equation on a flat substrate~\cite{Navascues1981,Widom1995}.  Equation  (\ref{eq:Br23}) also represents the force balance condition along the flat substrate.  The supersaturation or the undersaturation of vapor $\Delta\tilde{p}$ does not affect the boundary condition and only affects the capillary meniscus $f(z)$ from Eq.~(\ref{eq:Br9}).

\section{\label{sec:sec3}Line tension effect on capillary bridges}

In order to study the line tension effect on the capillary bridge, we will restrict ourselves to the simplest axisymmetric geometry where the upper and the lower substrate have exactly the same axisymmetric shape and the same wetting property so that the line tensions and Young's contact angles of two substrates are the same  ($\tau_{1}=\tau_{2}=\tau$, $\theta_{\rm Y1}=\theta_{\rm Y2}=\theta_{\rm Y}$ and $\theta_{1}=\theta_{2}=\theta$).

In general, the liquid-vapor meniscus $f(z)$ would be determined from the numerical solution of the Euler-Lagrange equation  (\ref{eq:Br8})~\cite{Erle1971,Orr1975}.  Although many analytical approximations such as the circular or the toroidal approximation~\cite{Fisher1927,Tselishchev2003,Butt2009} and others~\cite{Pepin2000,Lian2016,Kruyt2017} have been proposed, their use is restricted to special morphology~\cite{Fisher1927,Tselishchev2003,Butt2009} when the neck width of bridge is much wider than the narrow gap between two substrates.

In order to avoid numerical uncertainty due to the numerical solution of Eq.~(\ref{eq:Br8}) or the approximate expression for the meniscus $f(z)$, we will only consider the bridge at the liquid-vapor equilibrium when $\Delta\tilde{p}=0$ in Eq.~(\ref{eq:Br8}).  Even at the liquid-vapor equilibrium, the heterogeneous nucleation and capillary condensation can occur, and the liquid-vapor meniscus of the capillary bridge is the catenary~\cite{Orr1975,Malijevsky2015,Eriksson1995}
\begin{equation}
f\left(z\right)=w \cosh\left(\frac{z}{w}\right),
\label{eq:Br24}
\end{equation}
where $w$ is the neck width (radius) at $z=0$ where we choose the origin of the $z$ axis since the configuration of the two substrates is symmetric about $z=0$.  In the next subsections, we will use Eq.~(\ref{eq:Br24}) for several typical shapes $a_{1}(z)$ and $a_{2}(z)$ to study the interplay of line tension and geometry in the stability of the liquid bridge.  We start from the simplest and well-documented flat geometry formulated in section IIC.  Then, we will proceeds to the convex and the concave geometry in section IIA and B.

\subsection{Flat plate-plate geometry }

\begin{figure}[htbp]
\begin{center}
\includegraphics[width=1.0\linewidth]{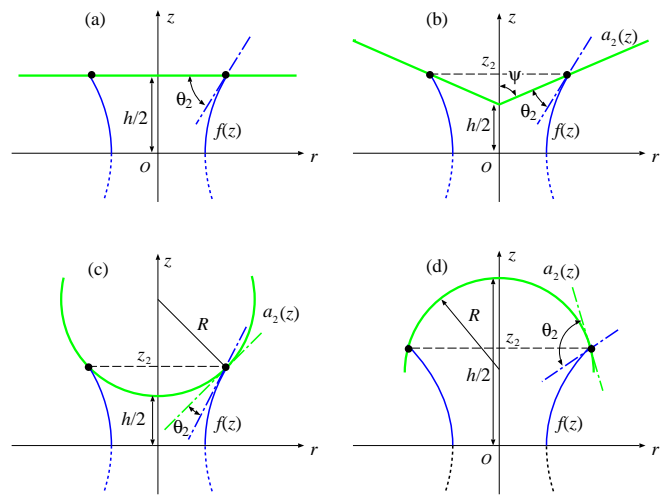}
\caption{
Various axisymmetric (a) flat plate-plate, (b) convex cone-cone, (c) convex sphere-sphere and (d) concave cap-cap geometry to study the line tension effect on the catenary capillary bridge in Eq.~(\ref{eq:Br24}).
 }
\label{fig:Br4}
\end{center}
\end{figure}

To begin with, we consider the catenary liquid bridge in Eq.~(\ref{eq:Br24}) formed between two identical flat and horizontal plates (Fig.~\ref{fig:Br4}(a)) separated by a gap height $h$. The modified Young's equation in Eq.(\ref{eq:Br21}) is written as
\begin{equation}
\cos\theta_{\rm Y}=\tanh \alpha + \hat{\tau}\frac{\alpha}{\cosh\alpha},
\label{eq:Br25}
\end{equation}
and the equilibrium contact angle in Eq.~(\ref{eq:Br22}) at the substrate is given by
\begin{equation}
\cos\theta=\tanh\alpha,
\label{eq:Br26}
\end{equation}
where 
\begin{equation}
      \alpha=\frac{h}{2w}
\label{eq:Br27}
\end{equation}
is the gap height relative to the bridge neck width $2w$, and
\begin{equation}
      \hat{\tau}=\frac{2\tilde{\tau}}{h}=\frac{2\tau}{h\sigma_{\rm lv}}
 \label{eq:Br28}
\end{equation}
is the non-dimensional line tension.  Suppose $\tau\sim 10^{-9}$ N (typical size)~\cite{Law2017}, and $\sigma_{\rm lv}\sim 2\times 10^{-2}$ Nm$^{-1}$ (typical alcohol), the non dimensional line tension becomes as large as $\hat{\tau}\sim 0.1$ for a nanoscale gap height $h\sim 50$ nm.

Figure~\ref{fig:Br5} shows the cosine of Young's contact angle $\theta_{\rm Y}$ versus the scaled bridge height $\alpha$ defined by Eq.~(\ref{eq:Br27}) calculated from Eq.~(\ref{eq:Br25}).  The morphology of the bridge $\alpha$ and the equilibrium contact angle $\theta$, which is in fact $\theta_{\rm Y}$ for $\hat{\tau}=0$ in Fig.~\ref{fig:Br5}, are determined from Young's contact angle $\theta_{\rm Y}$ as indicated by the up and the down arrow in Fig.~\ref{fig:Br5}.  The neck width $w$ of the bridge is determined from the gap height $h$ of the substrate through $\alpha$, which is determined from the wettability of substrate $\theta_{\rm Y}$.  

When the line tension is positive ($\hat{\tau}>0$), the contact line shrinks to avoid the increase of the free energy of the positive line tension.  Then the equilibrium contact angle $\theta$ would be larger than Young's contact angle $\theta_{\rm Y}$ so that $\cos\theta_{\rm Y}>\cos\theta>0$.  The substrate will be {\it less hydrophilic} when the line tension is positive.  For example, when $\cos\theta_{\rm Y}\sim 0.75$ and $\hat{\tau}=0.3$ in Fig.~\ref{fig:Br5},  the cosine of the equilibrium contact angle will be smaller $\cos\theta\sim 0.55$ (down arrow in Fig.~\ref{fig:Br5}).  In contrast, the negative line tension will make the substrate more hydrophilic so that the contact line expands and $0<\cos\theta_{\rm Y}<\cos\theta$ (up arrow in Fig.~\ref{fig:Br5}).  

\begin{figure}[htbp]
\begin{center}
\includegraphics[width=0.8\linewidth]{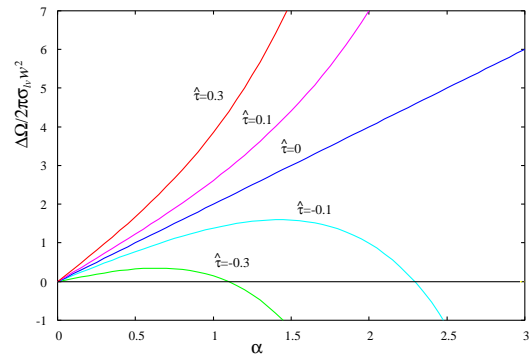}
\caption{
The grand potential $\Delta\Omega/2\pi\sigma_{\rm lv}w^{2}$ in Eq.~(\ref{eq:Br33}) as a function of the scaled gap height $\alpha$.
}
\label{fig:Br7}
\end{center}
\end{figure}

\begin{figure}[htbp]
\begin{center}
\includegraphics[width=0.9\linewidth]{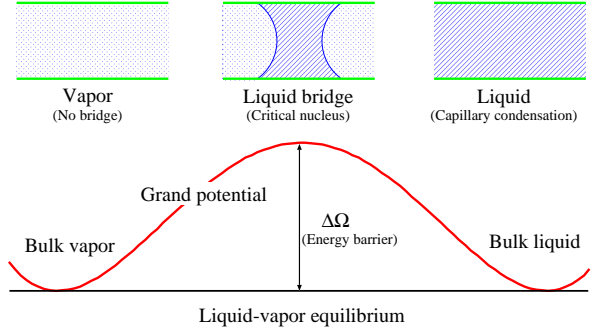}
\caption{
The schematic diagram of capillary condensation at the liquid-vapor equilibrium.  The grand potential $\Delta\Omega/2\pi\sigma_{\rm lv}w^{2}$ in Eq.~(\ref{eq:Br33}) corresponds to the energy barrier of heterogeneous nucleation of capillary condensation from vapor to liquid phase.
}
\label{fig:Br8}
\end{center}
\end{figure}

When the line tension is positive $\hat{\tau}>0$, there exists a maximum $\alpha_{\rm max}$ (a minimum $w$ for fixed $h$ or a maximum $h$ for fixed $w$) determined from 
\begin{equation}
\alpha_{\rm max}+\ln\alpha_{\rm max}=-\ln \hat{\tau},
\label{eq:Br29}
\end{equation}
which is derived from Eq.~(\ref{eq:Br25}) by setting $\cos\theta_{\rm Y}=1$ or $\theta_{\rm Y}=0^{\circ}$.  The catenary bridge cannot exist when $\cos\theta_{\rm Y}>1$ or $\alpha>\alpha_{\rm max}$ because the equilibrium contact angle $\theta$ cannot be realized for any realistic substrate with $\cos\theta_{\rm Y}\leq 1$.   In Fig.~{\ref{fig:Br6}, we show the maximum $\alpha_{\rm max}$ as a function of the scaled line tension $\hat{\tau}>0$.  When the line tension becomes larger, the maximum $\alpha_{\rm max}$ becomes smaller so that the existence of capillary bridge will be restricted.

\begin{figure}[htbp]
\begin{center}
\includegraphics[width=0.8\linewidth]{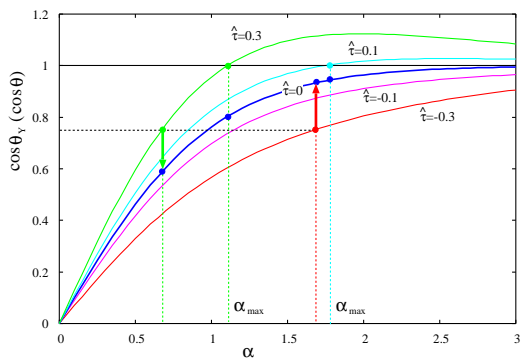}
\caption{
The cosine of Young's contact angle ($\cos\theta_{\rm Y}$) versus $\alpha$ for various sign and sizes of line tension $\hat{\tau}$ calculated from Eq.~(\ref{eq:Br25}).  When $\hat{\tau}$ is positive and $\alpha$ is larger than the maximum value $\alpha_{\rm max}$, the bridge cannot exist for any realistic substrate with  $\cos\theta_{\rm Y}\leq 1$.  The curve for $\hat{\tau}=0$ represents the cosine of the equilibrium contact angle $\cos\theta$ in Eq.~\ref{eq:Br26}, which will be actually realized. 
 }
\label{fig:Br5}
\end{center}
\end{figure}

\begin{figure}[htbp]
\begin{center}
\includegraphics[width=0.8\linewidth]{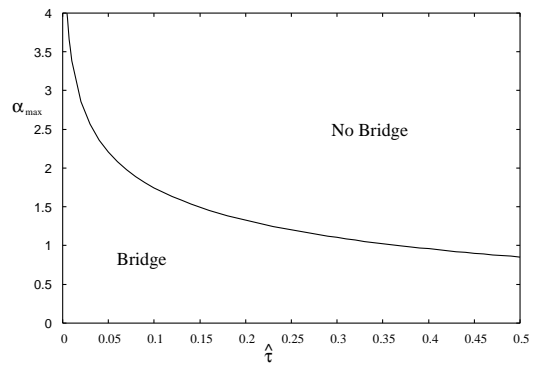}
\caption{
The maximum $\alpha_{\rm max}$ versus scaled positive line tension $\hat{\tau}$.  The height $h$ or the width $w$ of the bridge is bounded ($\alpha=h/2w<\alpha_{\rm max}$) when the line tension is positive.
 }
\label{fig:Br6}
\end{center}
\end{figure}

Since the shape of meniscus $f(z)$ and those of two substrates $a_{1}(z)$ and $a_{2}(z)$ are analytical functions, the free energy in Eqs.~(\ref{eq:Br1})-(\ref{eq:Br6}) can be analytically calculated~\cite{Eriksson1995}.  They are written as
\begin{eqnarray}
\Delta\Omega_{\rm lv} &=& \pi\sigma_{\rm lv}w^2\left[\sinh 2\alpha+2\alpha\right],
\label{eq:Br30}
\\
\Delta\Omega_{\rm sl}
&=& -\pi\sigma_{\rm lv}\cos\theta_{\rm Y}w^2\left[\cosh 2\alpha+1\right],
\label{eq:Br31}
\\
\Delta\Omega_{\rm slv} &=& 4\pi\sigma_{\rm lv}\tilde{\tau}\cosh \alpha,
\label{eq:Br32}
\end{eqnarray}
which give the total grand potential $\Delta\Omega=\Delta\Omega_{\rm lv}+\Delta\Omega_{\rm sl}+\Delta\Omega_{\rm slv}$ written as
\begin{equation}
\frac{\Delta\Omega}{2\pi\sigma_{\rm lv}w^{2}}=\alpha\left(1+2\hat{\tau}\cosh\alpha\right).
\label{eq:Br33}
\end{equation}
In Fig.~\ref{fig:Br7}, we show the total grand potential $\Delta\Omega$ as a function of $\alpha$ for various values of line tension $\hat{\tau}$. Apparently, the energy is mostly positive since it corresponds to the free energy barrier of heterogeneous nucleation of capillary condensation~\cite{Iwamatsu1996,Restagno2000,Desgranges2017,Kim2018} and the capillary bridge corresponds to the critical nucleus (Fig}.~\ref{fig:Br8}). The total grand potential becomes negative $\Delta\Omega<0$ only when the line tension is negative ($\hat{\tau}<0$) and $\alpha$ satisfies 
\begin{equation}
\alpha > {\rm arccosh}\frac{-1}{2\hat{\tau}}.
\label{eq:Br34}
\end{equation}

In Fig.~\ref{fig:Br8}, we show the schematic diagram of capillary bridge formation and subsequent capillary condensation.  The grand potential $\Delta\Omega/2\pi\sigma_{\rm lv}w^{2}$ in Eq.~(\ref{eq:Br33}) corresponds to the energy barrier of capillary condensation.  The capillary bridge is the critical nucleus of heterogeneous nucleation, and the capillary condensation occurs through the usual thermally activated process.  It cannot occur if $\cos\theta_{\rm Y}>1$ or $\alpha>\alpha_{\rm max}$ when the line tension is positive and large as shown in Figs.~\ref{fig:Br5} and \ref{fig:Br6} as the capillary bridge cannot exist.  The capillary condensation occurs only through the capillary bridge with $\alpha<\alpha_{\rm max}$ when the gap height $h$ is low relative to the bridge width $w$.  When the line tension is large and negative, this free energy can be negative as the length of contact line (or $\alpha$) becomes large and the line tension contribution $\Delta\Omega_{\rm slv}$ becomes dominant.  Then the capillary condensation can occur spontaneously via the capillary bridge, which satisfies the condition in Eq.~(\ref{eq:Br34}).

As has been noted in the previous section, the line tension cannot contribute directly to the normal capillary force parallel to the $z$ axis.  The contribution of line tension to the total grand potential (free energy) is featureless unless the line tension is large negative.  However, the line tension severely restrict the existence of the bridge as shown in Figs.~\ref{fig:Br5} and \ref{fig:Br6} when the line tension is positive because the low contact angle around the complete wetting of vanishing equilibrium contact angle $\theta=0$ cannot be realized.  In the next section, we will explore the bridge formation for the convex and the concave geometry.

\subsection{Convex cone-cone geometry }

Next, we consider the liquid bridge formed between two identical conical substrates (Fig.~\ref{fig:Br4}(b)) separated by the gap height $h$, which is the simplest model of AFM probe~\cite{Dutka2007,Dutka2010,Butt2009,Tselishchev2003}.   The functional forms $a_{1}(z)$ and $a_{2}(z)$ are given by
\begin{eqnarray}
a_{1}(z) &=& -\left(z+\frac{h}{2}\right)\tan\psi,\;\;\;z\leq -\frac{h}{2}
\label{eq:Br35}
\\
a_{2}(z) &=& \left(z-\frac{h}{2}\right)\tan\psi,\;\;\;z\geq \frac{h}{2}
\label{eq:Br36}
\end{eqnarray}
where $\psi$ is the half of the opening angle of the cone.

Since the bridge touches the substrate at $z_{2}$ (and $z_{1}=-z_{2}$), the contact condition $f\left(z_{2}\right)=a_{2}\left(z_{2}\right)$ given by
\begin{equation}
\cosh\alpha=\left(\alpha-\frac{h}{2w}\right)\tan\psi,
\label{eq:Br37}
\end{equation}
where
\begin{equation}
      \alpha=\frac{z_{2}}{w}
\label{eq:Br38}
\end{equation}
leads to
\begin{equation}
w=\frac{h/2}{\alpha-\cosh\alpha\cot\psi}.
\label{eq:Br39}
\end{equation}

\begin{figure}[htbp]
\begin{center}
\includegraphics[width=0.8\linewidth]{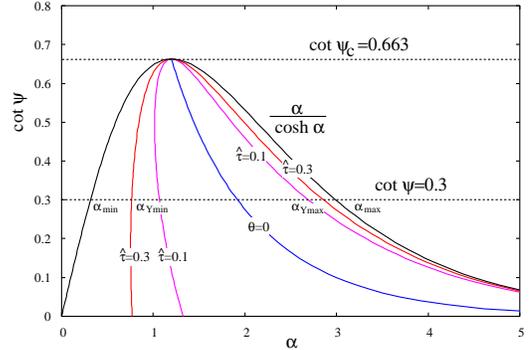}
\caption{
The phase diagram in variables $\left(\alpha, \cot\psi\right)$ showing the possible region of the existence of bridge. The bridge can exist only in the region enclosed by two curves $\alpha_{\rm min}$ and $\alpha_{\rm max}$ purely from the geometrical constraint.  When the line tension is positive $\hat{\tau}>0$, the region around the vanishing contact angle $\theta=0$ enclosed by $\alpha_{\rm Ymin}$ and $\alpha_{\rm Ymax}$ is further excluded and the bridge can exist only in two narrow regions between $\alpha_{\rm min}\leq \alpha \leq \alpha_{\rm Ymin}$ and $\alpha_{\rm Ymax}\leq\alpha\leq\alpha_{\rm max}$. 
 }
\label{fig:Br9}
\end{center}
\end{figure}

Since the neck width must be positive ($w>0$), the opening angle  $\psi$ of the cone must satisfy
\begin{equation}
\frac{\alpha}{\cosh\alpha}\ge \cot\psi
\label{eq:Br40}
\end{equation}
from Eq.~(\ref{eq:Br39}).  In Fig.~\ref{fig:Br9} we show the left-hand side of Eq.~(\ref{eq:Br40}) which has a maximum, where the neck-width $w$ of the bridge diverges ($w\rightarrow\infty$).   Therefore, the opening angle must be obtuse and larger than the critical angle $\psi_{\rm c}$, which can be easily obtained numerically at the maximum when $\cot\psi_{\rm c}=0.663$ (Fig.~\ref{fig:Br9}), and is given by
\begin{equation}
\psi_{\rm c}=56.5^{\circ}.
\label{eq:Br41}
\end{equation}
When the opening angle $\psi$ is larger than the critical angle $\psi_{\rm c}$, the position of the contact point $\alpha$ (or $z_{2}$) is limited by the inequality (\ref{eq:Br40}), and should stay within the range $\alpha_{\rm min}\leq\alpha\leq\alpha_{\rm max}$, where $\alpha_{\rm min}$ and $\alpha_{\rm max}$ are determined from
\begin{equation}
\frac{\alpha}{\cosh\alpha}=\cot\psi
\label{eq:Br42}
\end{equation}
and are shown in Fig.~\ref{fig:Br9}.  Therefore, the catenary capillary bridge in Eq.~(\ref{eq:Br24}) can exist only within the region enclosed by $\alpha_{\rm min}$ and $\alpha_{\rm max}$ in Fig.~\ref{fig:Br9} due to the geometrical constraint.

A similar concept of geometrically imposed stability bounds was argued by Finn~\cite{Finn1986} for more than two decades ago.  He used a refined mathematical formulation and discussed the stability of sessile droplet on a flat horizontal plane under the influence of the gravity.  The stability is controlled by the Bond number~\cite{Finn1986} and, therefore, by the gravity.  In our case, the geometrical constraint is due to the two walls, which sandwiches the droplet.  Therefore, the stability bounds come not from the gravity but from the incompatibility of contact angles at the two walls.

As far as the opening angle is larger than the critical angle ($\psi>\psi_{\rm c}$ or $\cot\psi<\cot\psi_{\rm c}$), the capillary bridge can exist within the range $\alpha_{\rm min}\leq\alpha\leq\alpha_{\rm max}$ shown in Fig.~\ref{fig:Br9}.  When the line tension is positive, the area around the equilibrium contact angle $\theta=0$ will be further excluded because the small contact angle around the complete wetting with $\theta=0$ can not be realized when $\hat{\tau}>0$ from the modified Young's equation in Eq.~(\ref{eq:Br26}).

In fact, the modified Young's equation (\ref{eq:Br26}) on the conical surface in Eq.~(\ref{eq:Br36}) is written as
\begin{equation}
\cos\theta_{\rm Y}=\frac{\cos\psi+\sin\psi\sinh\alpha}{\cosh\alpha} + \hat{\tau}\frac{\alpha}{\cosh\alpha}
\label{eq:Br43}
\end{equation}
and the equilibrium contact angle $\theta$ at the substrate is given by
\begin{equation}
\cos\theta=\frac{\cos\psi+\sin\psi\sinh\alpha}{\cosh\alpha},
\label{eq:Br44}
\end{equation}
from Eq.~(\ref{eq:Br11}), where the scaled line tension $\hat{\tau}$ is defined by Eq.~(\ref{eq:Br28}).

\begin{figure}[htbp]
\begin{center}
\includegraphics[width=0.85\linewidth]{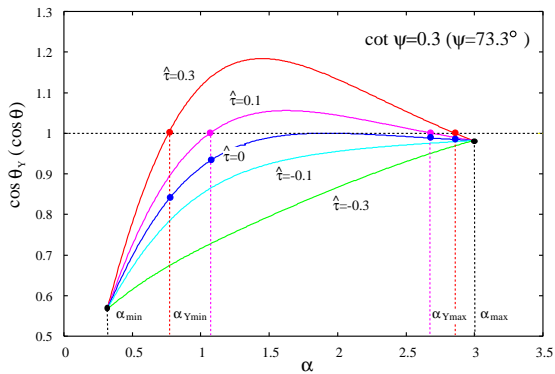}
\caption{
The cosine of Young's contact angle $\theta_{\rm Y}$ ($\cos\theta_{\rm Y}$) versus $\alpha$, which becomes larger than 1 ($\cos\theta_{\rm Y}>1$) in the interval $\alpha_{\rm Ymin}\leq\alpha\leq\alpha_{\rm Ymax}$ when the line tension is positive ($\hat{\tau}>0$).   Then, the bridge cannot exist for any realistic substrate with  $\cos\theta_{\rm Y}\leq 1$ and, the capillary bridge can exist only in two narrow intervals between $\alpha_{\rm min}\leq\alpha\leq\alpha_{\rm Ymin}$ and $\alpha_{\rm Ymax}\leq\alpha\leq\alpha_{\rm max}$ shown in Fig.~\ref{fig:Br9}.
 }
\label{fig:Br10}
\end{center}
\end{figure}

Figure~\ref{fig:Br10} shows the cosine of Young's contact angle $\cos\theta_{\rm Y}$ as a function of $\alpha$.  Since the cosine is always positive, the substrate must be hydorophilic to form a catenary capillary bridge in Eq.~(\ref{eq:Br24}).  When, the line tension is positive, the cosine of Young's contact angle $\theta_{\rm Y}$ can be larger than 1 ($\cos\theta_{\rm Y}>1$) as shown in Fig.~\ref{fig:Br10}, which means that the contact angle $\theta$ near the complete wetting ($\theta=0$) cannot be realized for any realistic substrate with  $\cos\theta_{\rm Y}\leq 1$.  The substrate becomes less hydrophilic $\theta>\theta_{\rm Y}$ or $\cos\theta<\cos\theta_{\rm Y}$ by the effect of positive line tension.  Then, the interval between $\alpha_{\rm Ymin}$ and $\alpha_{\rm Ymax}$ around $\theta=0$ is not accessible, where $\alpha_{\rm Ymin}$ and $\alpha_{\rm Ymax}$ are the solutions of $\cos\theta_{\rm Y}=1$ of Eq.~(\ref{eq:Br43}).  The capillary bridge can exist only in two narrow intervals between $\alpha_{\rm min}<\alpha<\alpha_{Ymin}$ and $\alpha_{\rm Ymax}<\alpha<\alpha_{\rm max}$ shown in Fig.~\ref{fig:Br10}.  The equilibrium contact angle cannot reach zero and has a minimum angle, and the existence of capillary bridge will be further restricted when the line tension is positive.

Again, it is possible to write down the analytical formula for the grand potential for the cone-cone geometry, which becomes exactly the same form as in Eq.~(\ref{eq:Br33}).  Since the grand potential is featureless as it corresponds to the energy barrier of heterogeneous nucleation, we will not consider the grand potential (free energy) in the following discussion.

\subsection{Convex sphere-sphere geometry }

As the third example, we consider the liquid bridge formed between two identical spherical substrates of radius $R$ (Fig.~\ref{fig:Br4}(c)) separated by the gap height $h$.   This geometry is a typical model of various problems of science and engineering, and has been the subject of continuous interest of many scientists and engineers for many years~\cite{Haines1925,Fisher1927,Pepin2000,Kruyt2017,Erle1971,Orr1975,Lian2016,Butt2009,Asay2010,Laube2015,Dobbs1992a,Bauer2000,Malijevsky2015,MacDowell2018}.    

The functional forms $a_{1}(z)$ and $a_{2}(z)$ of the two spherical substrates are given by
\begin{eqnarray}
a_{1}(z) &=& \sqrt{R^2-\left(z+R+\frac{h}{2}\right)^2},\;\;\;z\leq -\frac{h}{2},
\label{eq:Br45}
\\
a_{2}(z) &=& \sqrt{R^2-\left(z-R-\frac{h}{2}\right)^2},\;\;\;z\geq \frac{h}{2},
\label{eq:Br46}
\end{eqnarray}
where $R$ is the radius of two spheres separated by $h$ (Fig.~\ref{fig:Br4}(c)).

\begin{figure}[htbp]
\begin{center}
\includegraphics[width=0.85\linewidth]{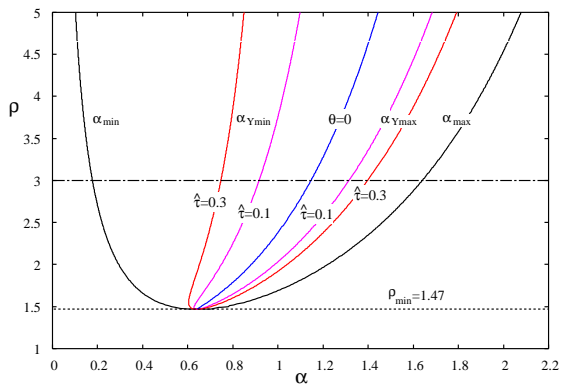}
\caption{
The phase diagram in variables $\left(\alpha, \rho\right)$ showing the possible region of the existence of catenary bridge.  The bridge can exist only in the region enclosed by two curves $\alpha_{\rm min}$ and $\alpha_{\rm max}$.  When the line tension is positive $\hat{\tau}>0$, the region around the vanishing contact angle $\theta=0$ enclosed by $\alpha_{\rm Ymin}$ and $\alpha_{\rm Ymax}$ is further excluded.  Then, the bridge can exist only in two narrow regions between $\alpha_{\rm min}\leq \alpha \leq \alpha_{\rm Ymin}$ and $\alpha_{\rm Ymax}\leq\alpha\leq\alpha_{\rm max}$. 
 }
\label{fig:Br11}
\end{center}
\end{figure}

In this case, the contact condition $f\left(z_{2}\right)=a_{2}\left(z_{2}\right)$ leads to 
\begin{equation}
w=\frac{h/2}{\alpha-\rho+\sqrt{\rho^2-\cosh^{2}\alpha}},
\label{eq:Br47}
\end{equation}
where
\begin{equation}
\rho=\frac{R}{w}
\label{eq:Br48}
\end{equation}
is the scaled radius of the spherical substrate and $\alpha$ is defined by Eq.~(\ref{eq:Br38}).  Since the neck width must be positive $w>0$, the position of contact point $\alpha$ and the scaled radius $\rho$ of the sphere satisfy
\begin{equation}
\alpha-\rho+\sqrt{\rho^2-\cosh^{2}\alpha}\ge 0
\label{eq:Br49}
\end{equation}
from Eq.~(\ref{eq:Br47}).  Since the left-hand side of Eq.~(\ref{eq:Br47}) is a convex function of $\alpha$ and has a maximum at $\alpha=\alpha_{c}$ given by
\begin{equation}
\cosh^{2}\alpha_{c}=\rho,
\label{eq:Br50}
\end{equation}
the inequality (\ref{eq:Br49}) will be satisfied as far as the maximum of Eq.~(\ref{eq:Br47}) at $\alpha_{c}$ is positive. This condition is written as
\begin{equation}
{\rm arccosh}\sqrt{\rho}-\rho+\sqrt{\rho\left(\rho-1\right)}\geq 0
\label{eq:Br51}
\end{equation}
from Eq.~(\ref{eq:Br49}).  Equation (\ref{eq:Br51}) can be satisfied when $\rho\geq\rho_{\rm min}$, where
\begin{equation}
\rho_{\rm min}=1.47
\label{eq:Br52}
\end{equation}
at
\begin{equation}
\alpha_{c}=0.639.
\label{eq:Br53}
\end{equation}
When the radius of the sphere is larger than this minimum size ($\rho>\rho_{\rm min}$), the capillary bridge can exist only when $\alpha_{\rm min}\leq\alpha\leq\alpha_{\rm max}$, where $\alpha_{\rm min}$ and $\alpha_{\rm max}$ are the solution of
\begin{equation}
\alpha-\rho+\sqrt{\rho^2-\cosh^{2}\alpha}= 0,\;\;\;\rho\ge\rho_{\rm min},
\label{eq:Br54}
\end{equation}
from Eq.~(\ref{eq:Br49}) and are shown in Fig.~\ref{fig:Br11}.

The modified Young's equation in Eq.(\ref{eq:Br10}) for the sphere-sphere geometry is given by
\begin{eqnarray}
\cos\theta_{\rm Y} &=& \frac{\cosh\alpha+\sinh\alpha\sqrt{\rho^2-\cosh^{2}\alpha}}{\rho\cosh\alpha}
\nonumber \\
&+& \hat{\tau}\left(\alpha-\rho+\sqrt{\rho^2-\cosh^{2}\alpha}\right)\frac{\sqrt{\rho^2-\cosh^{2}\alpha}}{\rho\cosh\alpha}, 
\nonumber \\
\label{eq:Br55}
\end{eqnarray}
where $\hat{\tau}$ is defined by Eq.~(\ref{eq:Br28}), and the equilibrium contact angle $\theta$ at the substrate is given by
\begin{equation}
\cos\theta=\frac{\cosh\alpha+\sinh\alpha\sqrt{\rho^2-\cosh^{2}\alpha}}{\rho\cosh\alpha},
\label{eq:Br56}
\end{equation}
from Eq.~(\ref{eq:Br11}), which is Eq.~(\ref{eq:Br55}) when $\hat{\tau}=0$.

In this sphere-sphere geometry, the cosine of Young's contact angle $\cos\theta_{\rm Y}$ calculated from Eq.~(\ref{eq:Br55}) shows a maximum similar to that in Fig.~\ref{fig:Br10} and becomes larger than 1 when the line tension is positive.  Since $\cos\theta_{\rm Y}>\cos\theta$, the substrate is less hydrophilic ($\theta_{\rm Y}<\theta$) when the line tension is positive. Then, the region around $\theta=0$ in Fig.~\ref{fig:Br11} will be excluded when  the line tension is positive ($\hat{\tau}>0$).  Similar to the case of the cone-cone geometry, the capillary bridge can exist only in two narrow interval between $\alpha_{\rm min}<\alpha<\alpha_{\rm Ymin}$ and $\alpha_{\rm Ymax}<\alpha<\alpha_{\rm max}$ shown in Fig.~\ref{fig:Br11}, where $\alpha_{\rm Ymin}$ and $\alpha_{\rm Ymax}$ are the solutions of $\cos\theta_{\rm Y}=1$ in Eq.~(\ref{eq:Br55}).

\begin{figure}[htbp]
\begin{center}
\includegraphics[width=0.85\linewidth]{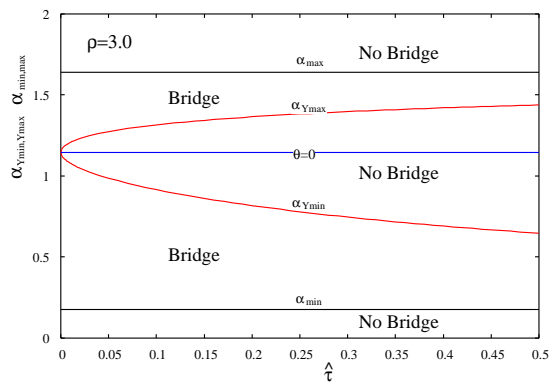}
\caption{
The two boundaries $\alpha_{\rm Ymin}$ and $\alpha_{\rm Ymax}$ as functions of the scaled line tension $\hat{\tau}$ when $\rho=3.0$ indicated by the horizontal line in Fig.~\ref{fig:Br11}..   The horizontal lines represent $\alpha_{\rm min}$, $\alpha_{\rm max}$ and the line corresponds to the equilibrium contact angle $\theta=0$, which do not depend on the size of line tension $\hat{\tau}$.
 }
\label{fig:Br12}
\end{center}
\end{figure}

In figure~\ref{fig:Br12} we show the line-tension dependence of the two boundaries $\alpha_{\rm Ymin}$ and $\alpha_{\rm Ymax}$ as well as $\alpha_{\rm min}$, $\alpha_{\rm max}$ and $\alpha$ that corresponds to the vanishing equilibrium contact angle $\theta=0$ when $\rho=3.0$ indicated by the horizontal line in Fig.~\ref{fig:Br11}.   The bridge cannot exist outside the two boundaries $\alpha_{\rm min}$ and $\alpha_{\rm max}$.  In addition, the region around the complete wetting state with $\theta=0$ is excluded when the line tension is positive ($\hat{\tau}>0$).  Two intervals between $\alpha_{\rm min}<\alpha<\alpha_{\rm Ymin}$ and  $\alpha_{\rm Ymax}<\alpha<\alpha_{\rm max}$ where bridge can exist become narrower as the line tension becomes larger.

\subsection{Concave cap-cap geometry }

Finally, we consider the case when the substrate is concave.  As a simplest example, we consider the catenary capillary bridge formed between two half-spherical cap-shaped substrates.  This problem would be relevant to some biological problems such as the wet adhesion of insect legs~\cite{Federle2002,Su2009}.

\begin{figure}[htbp]
\begin{center}
\includegraphics[width=0.85\linewidth]{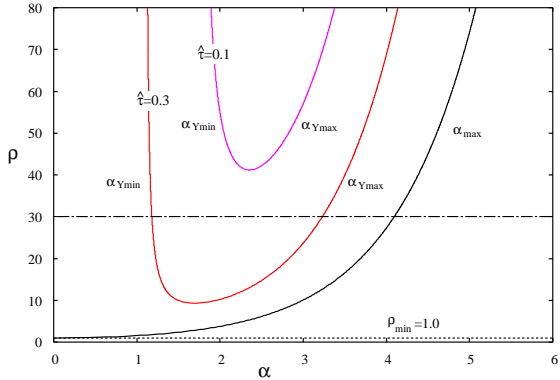}
\caption{
The phase diagram in variables $\left(\alpha, \rho\right)$ showing the possible region of existence of bridge.  The bridge can exist above the curve $\rho=\cosh\alpha_{\rm max}$ and $\rho>\rho_{\rm min}=1$.  When the line tension is positive $\hat{\tau}>0$ and the radius $\rho=R/w$ is large, the region around small contact angle $\theta=0$ enclosed by $\alpha_{\rm Ymin}$ and $\alpha_{\rm Ymax}$ is further excluded.  Then, the bridge can exist only in two regions between $0 \leq \alpha \leq \alpha_{\rm Ymin}$ and $\alpha_{\rm Ymax}\leq\alpha\leq\alpha_{\rm max}$
 }
\label{fig:Br13}
\end{center}
\end{figure}

The shapes of two substrates are given by
\begin{eqnarray}
a_{1}(z) &=& \sqrt{R^2-\left(z+R-\frac{h}{2}\right)^2},\;\;\;R-\frac{h}{2}\leq z\leq -\frac{h}{2}, \nonumber \\
\label{eq:Br57}
\\
a_{2}(z) &=& \sqrt{R^2-\left(z-R-\frac{h}{2}\right)^2},\;\;\; \frac{h}{2}-R\leq z\leq \frac{h}{2}. \nonumber \\
\label{eq:Br58}
\end{eqnarray}
In this case, the contact condition $f\left(z_{2}\right)=a_{2}\left(z_{2}\right)$ leads to 
\begin{equation}
w=\frac{h/2}{\alpha+\rho-\sqrt{\rho^2-\cosh^{2}\alpha}}
\label{eq:Br59}
\end{equation}
and the bridge can exist as far as
\begin{equation}
\rho\ge \cosh\alpha.
\label{eq:Br60}
\end{equation}
Given the radius of substrate $R$ or $\rho$, the maximum of $\alpha$ is determined from
\begin{equation}
\cosh\alpha_{\rm max}=\rho.
\label{eq:Br61}
\end{equation}
In Fig.~\ref{fig:Br13}, we show the curve $\rho=\cosh\alpha_{\rm max}$ in Eq.~(\ref{eq:Br61}).  The bride can exist above this curve from Eq.~(\ref{eq:Br60}).  Clearly there exists a lower bound for the scaled radius $\rho_{\rm min}=1$ below which the bridge cannot exist.  However, the bridge between concave substrates can exist in relatively wide area in Fig.~\ref{fig:Br13} compared to the bridges between convex substrates in Figs.~\ref{fig:Br9} and \ref{fig:Br11}.

The modified Young's equation for the cap-shaped concave substrate is written as
\begin{eqnarray}
\cos\theta_{\rm Y} &=& \frac{\cosh\alpha-\sinh\alpha\sqrt{\rho^2-\cosh^{2}\alpha}}{\rho\cosh\alpha}
\nonumber \\
&-& \hat{\tau}\left(\alpha+\rho-\sqrt{\rho^2-\cosh^{2}\alpha}\right)\frac{\sqrt{\rho^2-\cosh^{2}\alpha}}{\rho\cosh\alpha}, 
\nonumber \\
\label{eq:Br62}
\end{eqnarray}
from Eq.~(\ref{eq:Br17}) instead of Eq.~(\ref{eq:Br10}), and the equilibrium contact angle $\theta$ at the substrate is given by
\begin{equation}
\cos\theta=\frac{\cosh\alpha-\sinh\alpha\sqrt{\rho^2-\cosh^{2}\alpha}}{\rho\cosh\alpha}.
\label{eq:Br63}
\end{equation}
from Eq.~(\ref{eq:Br18}).  These results in Eqs.~(\ref{eq:Br62}) and (\ref{eq:Br63}) are similar to those of the convex sphere-sphere geometry in Eqs.~(\ref{eq:Br55}) and (\ref{eq:Br56}).

\begin{figure}[htbp]
\begin{center}
\includegraphics[width=0.85\linewidth]{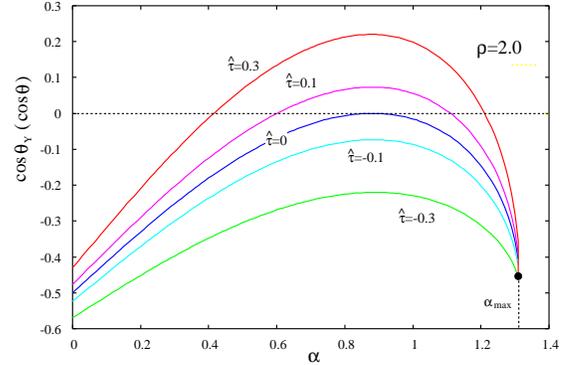}
\caption{
The cosine of Young's contact angle ($\cos\theta_{\rm Y}$) versus $\alpha$ when the radius of substrate is small $\rho=2.0$. The cosine is mostly negative ($\cos\theta_{\rm Y}<0$) indicating the hydrophobic substrate with the contact angle higher than $90^{\circ}$ which is necessary for bridge formation when the radius of the substrate $\rho$ is small.  Then, the capillary bridge can exist only on hydrophobic substrates. }
\label{fig:Br14}
\end{center}
\end{figure}

In Fig.~\ref{fig:Br14}, we show the cosine of Young's contact angle $\cos\theta_{\rm Y}$ as a function of $\alpha$ when the radius of the substrate is $\rho=2.0$.  Naturally, the cosine is mostly negative.  Therefore, the hydrophobic substrate is necessary for the bridge formation when the radius of the substrate $\rho$ is small.  Since the substrate will be less hydrophilic or more hydrophobic by positive line tensions, the hydrophilic substrate with $\cos\theta_{\rm Y}>0$ (e.g. the line $\hat{\tau}=0.3$ in Fig.~\ref{fig:Br14}) turn to effectively hydrophobic with $\cos\theta<0$ (the line $\hat{\tau}=0$ in Fig.~\ref{fig:Br14}).

\begin{figure}[htbp]
\begin{center}
\includegraphics[width=0.85\linewidth]{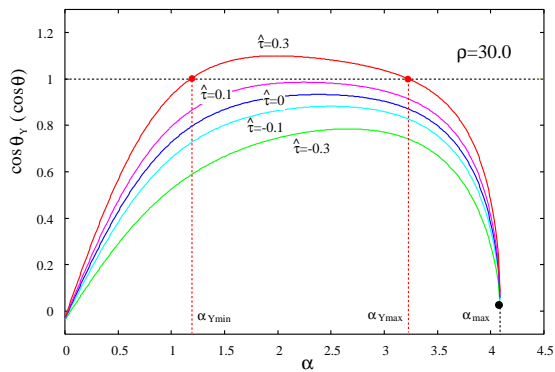}
\caption{
The cosine of Young's contact angle ($\cos\theta_{\rm Y}$) versus $\alpha$ when the radius of substrate is large $\rho=30.0$ indicated by the horizontal line in Fig.~\ref{fig:Br13}.  The cosine is positive and can be larger than 1 ($\cos\theta_{\rm Y}>1$) in the interval $\alpha_{\rm Ymin}\leq\alpha\leq\alpha_{\rm Ymax}$ when the line tension is positive and large ($\hat{\tau}=0.3$).  Then, the capillary bridge can exist only in two narrow intervals between $0 \leq\alpha\leq\alpha_{\rm Ymin}$ and $\alpha_{\rm Ymax}\leq\alpha\leq\alpha_{\rm max}$ shown in Fig.~\ref{fig:Br13}
 }
\label{fig:Br15}
\end{center}
\end{figure}

When the radius $\rho$ is large, two concave substrates become flatter so that the effect of line tension would become noticeable.   In fact, when the radius $\rho$ is large and the line tension is positive ($\hat{\tau}>0$), the cosine of Young's contact angle $\theta_{\rm Y}$ can be positive and larger than 1 ($\cos\theta_{\rm Y}>1$) as shown in Fig.~\ref{fig:Br15}.  The concave cap-shaped hydrophilic substrate can sustain the catenary bridge.  The interval between $\alpha_{\rm Ymin}$ and $\alpha_{\rm Ymax}$ around $\theta=0^{\circ}$ will not be accessible, where $\alpha_{\rm Ymin}$ and $\alpha_{\rm Ymax}$ are the solutions of $\cos\theta_{\rm Y}=1$ of Eq.~(\ref{eq:Br62}).  Then, the capillary bridge can exist only in two intervals between $0<\alpha<\alpha_{\rm Ymin}$ and $\alpha_{\rm Ymax}<\alpha<\alpha_{\rm max}$ shown in Fig.~\ref{fig:Br13}.  The equilibrium contact angle cannot reach zero and has a minimum angle (Fig.~\ref{fig:Br15}). The existence of capillary bridge with the equilibrium contact angle close to zero will be prohibited even for concave substrates when the radius $\rho$ is large and the line tension $\hat{\tau}$ is positive.   The story is similar to that for the convex spherical substrates.  However, the radius $\rho$ for the concave substrate must be much larger than that for the convex substrate (Figs.~\ref{fig:Br11} and \ref{fig:Br13}).

\begin{figure}[htbp]
\begin{center}
\includegraphics[width=0.95\linewidth]{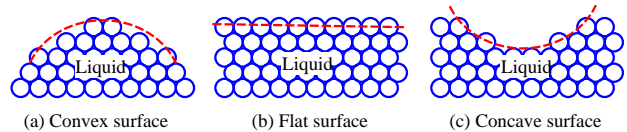}
\caption{
The surface tension of convex, flat and concave surfaces.  Since the number of broken bonds between neighboring atoms would be larger for the convex substrate and smaller for the concave substrate than that for the flat substrate, the surface tension would be higher for the convex substrate and lower for the concave substrate than that for the flat substrate.
 }
\label{fig:Br16}
\end{center}
\end{figure}

Recently, there appear the normal capillary force measurements of nanoscale bridges~\cite{Kwon2018,Kim2018}, which are interpreted by the size- or curvature-dependent liquid-vapor surface tension of nanoscale liquids using Tolman's formula~\cite{Tolman1949,Iwamatsu1994}.  As shown in Fig.~\ref{fig:Br16}, the atomic bonds between neighboring atoms are broken near the liquid-vapor interface, which results in an increase of surface energy called surface tension. Since the number of broken bonds between neighboring atoms would be larger for the convex substrate and smaller for the concave substrate than that for the flat substrate, the surface tension would be higher for the convex substrate and lower for the concave substrate than that for the flat substrate.  Therefore, the curvature correction which is represented by the Tolman's length can be negative for convex surface and positive for concave surface~\cite{Iwamatsu1994}. In fact, several model calculations~\cite{Iwamatsu1994,Kanduc2018} indicate both positive and negative Tolman's length. The issue of curvature-dependent surface tension for membranes and vesicles with elastic surface is also well know.~\cite{Helfrich1973,Seifert1997}.  In this article, however, we concentrate on the capillary bride with liquid surface.

However, the curvature of bridge surface is a combination of a positive and a negative curvature and, in particular, it vanishes for the catenary meniscus at the liquid-vapor equilibrium.  Then, Tolman's curvature correction to the surface tension would be very small or vanish. Then, the line tension would be another option to consider the nanoscale effect on liquid capillary bridges.  Of course, the real surface of liquid is not as sharp as Fig.~\ref{fig:Br16}. The real surface must be diffuse~\cite{Evans1979,Dellisola1995} which further contributes to the size or the curvature dependence of surface tension.  Therefore, the effect of size- or curvature-dependent liquid-vapor surface tension to the capillary bridge might not be negligible.  Of course, the line tension is not merely a phenomenological parameter but it needs detailed consideration and should be interpreted as an effective line tension which includes various nanoscale effects such as the disjoining pressure~\cite{Indekeu1994,Aveyard1999}, Tolman's correction to the surface tension~\cite{Law2017,Iwamatsu2018,Kanduc2018} and the adsorption to the substrate~\cite{Das2018,Tatyanenko2019}.  At the present stage, even the sign of line tension is unpredictable.~\cite{Law2017}  Further experimental and theoretical effort is certainly necessary to elucidate the nature of nanoscale capillary bridge with a positive and negative curvature.

\section{\label{sec:sec4} Conclusion}

In the present study, we considered the effect of line tension on the catenary capillary bridge between two identical substrates with convex, concave and flat geometry using the classical capillary theory.  The modified Young's equation, which takes into account the effect of line tension, is derived on a general axisymmetric curved surface from the variation of the grand potential.  Although the derived formula is used to study the effect of line tension only on flat, conical and spherical substrates, it is easy to extend our analysis to other geometry as well~\cite{Tselishchev2003,Butt2009,Wang2016}.  Since we did not consider the disjoining pressure or the surface potential, the adsorption layer around the substrates is not considered~\cite{Dobbs1992a,Bauer2000,Malijevsky2015,MacDowell2018}.  This problem can be partially circumvented by regarding the effective substrate which dresses the adsorption layer~\cite{Asay2010,Laube2015}.  Although only the catenary bridge at the liquid-vapor equilibrium is considered, the results will be applicable as far as the supersaturation or the undersaturation of vapor pressure is not too far from the liquid-vapor equilibrium.

It is clearly demonstrated that even without the effect of line tension, the existence of bridge is already restricted simply by geometrical constraint.  The modified Young's equation further restricts the bridge formation when the line tension is positive because the equilibrium contact angle which is necessary for the bridge formation cannot be achieved.  Then, the heterogeneous nucleation of capillary bridge formation and subsequent capillary condensation will not occur.  Although the line tension cannot contribute directly to the normal capillary force between two substrates, the line tension indirectly affects the force through modification of the equilibrium contact angle~\cite{Duriez2017}.  Therefore, the interpretation of normal capillary force of nanoscale bridges needs caution, and the contact angle determination combined with the normal capillary force measurement should be carefully conducted.

\begin{acknowledgments}
This work was conducted at the Department of Physics, Tokyo Metropolitan University (TMU) while one of the authors (MI) has been a visiting scientist .  The author is grateful to the Department of Physics (TMU), Professor Hiroyuki Mori and Professor Yutaka Okabe for their support and warm hospitality.  
\end{acknowledgments}

\appendix*
\section{}
The transversality condition for the grand potential in Eq.~(\ref{eq:Br1}) at $z_{2}$ on a convex substrate
\begin{equation}
\frac{\delta}{\delta z_{2}}\left[\frac{\delta\Delta\tilde{\Omega}}{\delta f_{2}}\right]=0,
\label{eq:Ap1}
\end{equation}
where $f_{2}=f\left(z_{2}\right)$, leads to~\cite{Hildebrand1992}
\begin{eqnarray}
&&\left[\left(f_{2}\sqrt{1+f_{2}'^{2}}-\cos\theta_{Y2}a_{2}\sqrt{1+a_{2}'^{2}}+\tilde{\tau}_{2}a_{2}'\right)\right]\delta z_{2} \nonumber \\
&+&\left[\frac{\partial F}{\partial f'}\right]_{z_{2}}\delta f_{2}-\int_{z_{1}}^{z_{2}}\left(\frac{d}{dz}\left(\frac{\partial F}{\partial f'}\right)-\frac{\partial F}{\partial f}\right)\delta f dz=0,
\nonumber \\
\label{eq:Ap2}
\end{eqnarray}
where we have used abbreviations $f_{2}'=f'\left(z_{2}\right)$, $a_{2}'=a_{2}'\left(z_{2}\right)$ and $a_{2}=a_{2}\left(z_{2}\right)$.  

The last term of Eq.~(\ref{eq:Ap2}) gives the Euler-Lagrange equation in Eq.~(\ref{eq:Br8}), which is explicitly written as in Eq.~(\ref{eq:Br9}) by using
\begin{eqnarray}
\frac{\partial F}{\partial f'} &=& \frac{ff'}{\sqrt{1+f'^{2}}}
\label{eq:Ap3}
\\
\frac{\partial F}{\partial f} &=& \frac{1}{\sqrt{1+f'^{2}}}+2\Delta\tilde{p}f
\label{eq:Ap4}
\end{eqnarray}
from Eq.~(\ref{eq:Br3}).

Due to the boundary condition $a_{2}=f_{2}$ at $z_{2}$, we have~\cite{Hildebrand1992}
\begin{equation}
\delta z_{2}=\frac{\delta f_{2}}{a_{2}'-f_{2}'}.
\label{eq:Ap5}
\end{equation}
Using Eqs.~(\ref{eq:Ap3}) and (\ref{eq:Ap5}), the first and the second term in Eq.~(\ref{eq:Ap2}) immediately leads to modified Young's equation for the convex substrate given by Eq.~(\ref{eq:Br10}) of the main text.



\end{document}